\documentclass[amsmath,amssymb,aps,pre]{revtex4-2}
\usepackage{graphicx} 
\usepackage{dcolumn}   
\usepackage{bm}        
\usepackage{amssymb}   
\usepackage{mathtools}
\usepackage{amsmath}
\usepackage{color}
\usepackage{dblfloatfix} 
\usepackage{hyperref}
\usepackage{epsf,epsfig}

\begin{document}

\title{Emergent learning in physical systems as feedback-based aging in a glassy landscape}
\author{Vidyesh Rao Anisetti,$^1$ Ananth Kandala,$^{2}$ J. M. Schwarz$^{1,3}$}

\affiliation{$^1$Physics Department, Syracuse University, Syracuse, NY 13244 USA\\
$^2$Department of Physics, University of Florida,  FL 32611-8440, USA\\
$^3$Indian Creek Farm, Ithaca, NY 14850 USA }
\date{\today}
\begin{abstract}
   By training linear physical networks to learn linear transformations, we discern how their physical properties evolve due to weight update rules. Our findings highlight a striking similarity between the learning behaviors of such networks and the processes of aging and memory formation in disordered and glassy systems. We show that the learning dynamics resembles an aging process, where the system relaxes in response to repeated application of the feedback boundary forces in presence of an input force, thus encoding a memory of the input-output relationship. With this relaxation comes an increase in the correlation length, which is indicated by the two-point correlation function for the components of the network. We also observe that the square root of the mean-squared error as a function of epoch takes on a non-exponential form, which is a typical feature of glassy systems. This physical interpretation  suggests that by encoding more detailed information into input and feedback boundary forces, the process of emergent learning can be rather ubiquitous and, thus, serve as a very early physical mechanism, from an evolutionary standpoint, for learning in biological systems.
\end{abstract}

\maketitle

\section{Introduction}
Given the prevalence of emergent behavior, physicists, computer scientists, and biologists have long asked whether or not some subset of emergent behavior results in the capacity of a system of many interacting components to learn, i.e., to have intelligence~\cite{Hopfield1982neural,Hillis1988intelligence}. While there has been much focus looking for emergent learning in brain-like systems, such as neuronal networks in biology or artificial neural networks in physics and computer science, recent research has demonstrated that simple physical systems, such as a spring network, have the potential to exhibit learning behavior similar to that of artificial neural networks~\cite{Anisetti2023,scellier2021deep,anisetti2022frequency,stern2021,kendall2020training,stern_review,murgan_stern_folding}. In this context, learning refers to the ability to modify the properties of a physical system by adjusting its learning degrees of freedom in order to more efficiently achieve some task.  For example, in a spring network, the spring stiffness and rest lengths represent the learning degrees of freedom, while the nodes of the springs correspond to the usual physical degrees of freedom. 

In these physical learning systems, once  input boundary nodes, output boundary nodes, and a cost function are all chosen, the learning process is composed of two steps: \begin{enumerate}
    \item {\it Signaling } : System's response to a given input is compared with the desired output and an update signal is sent which provides information on the necessary adjustments to each learning degree of freedom, so that the system's response aligns more closely with the desired output.
    \item  {\it Weight update} : Each learning degree of freedom, or weight, is updated in response to the update signal. This weight update should allow the system to perform gradient descent. 
\end{enumerate} 
The two steps are repeatedly applied to train the system to learn. 

The major challenge applying this algorithm is to find physical processes that implement the above two steps. While methods such as Equilibrium Propagation (EP)~\cite{scellier2021deep}, Multi-mechanism Learning (MmL)~\cite{Anisetti2023,anisetti2022frequency}, and Coupled Learning (CP)~\cite{stern2021} have made strides in addressing this challenge, they are not entirely physical in nature. In particular, the learning stages involved,  {\it Signaling} and {\it Weight update}, require the artificial modifications to the physical system. For instance, in EP and CL, to send the gradient information into the system, one needs to store the free state in some memory, which is not possible in typical systems such as spring networks or resistor networks. In our previous work unveiling MmL, we demonstrated that this issue of memory storage could be addressed by encoding the feedforward and feedback signal into two non-interfering physical quantities~\cite{Anisetti2023,anisetti2022frequency}. 

Despite this demonstration, however, a significant problem remains: we do not know of any physical process that can update the weights in the system. To physically implement weight updates, recent experimental efforts have resorted to using complex components such as transistors in the training of electrical networks~\cite{Dillavou2023,Dillavou2022}, and actuators and sensors in mechanical networks~\cite{lee2022mechanical}. Yet, the reliance on such intricate and varied tools introduces challenges in terms of scalability and robustness in these approaches. Here, we explore the central question: Do effects of the weight update procedure resemble any natural physical phenomena? The answer to such a question will point us in the direction of a fully physical learning system, weight update included. To begin to answer this question, we train linear physical networks and investigate how the physical properties of this system change, given the weight update rule.

Our manuscript consists of revisiting our MmL training procedure, as detailed in our prior  work~\cite{Anisetti2023,anisetti2022frequency}, in a general manner that emphasizes its physical plausibility. We then review the specifics of multi-mechanism learning, followed by details of what we measure as well as data generation and network generation. Results are then presented.  We conclude with a discussion of the impact of our results.  

\begin{figure*}[htb!]
    \centering  \includegraphics[width=0.99\textwidth]{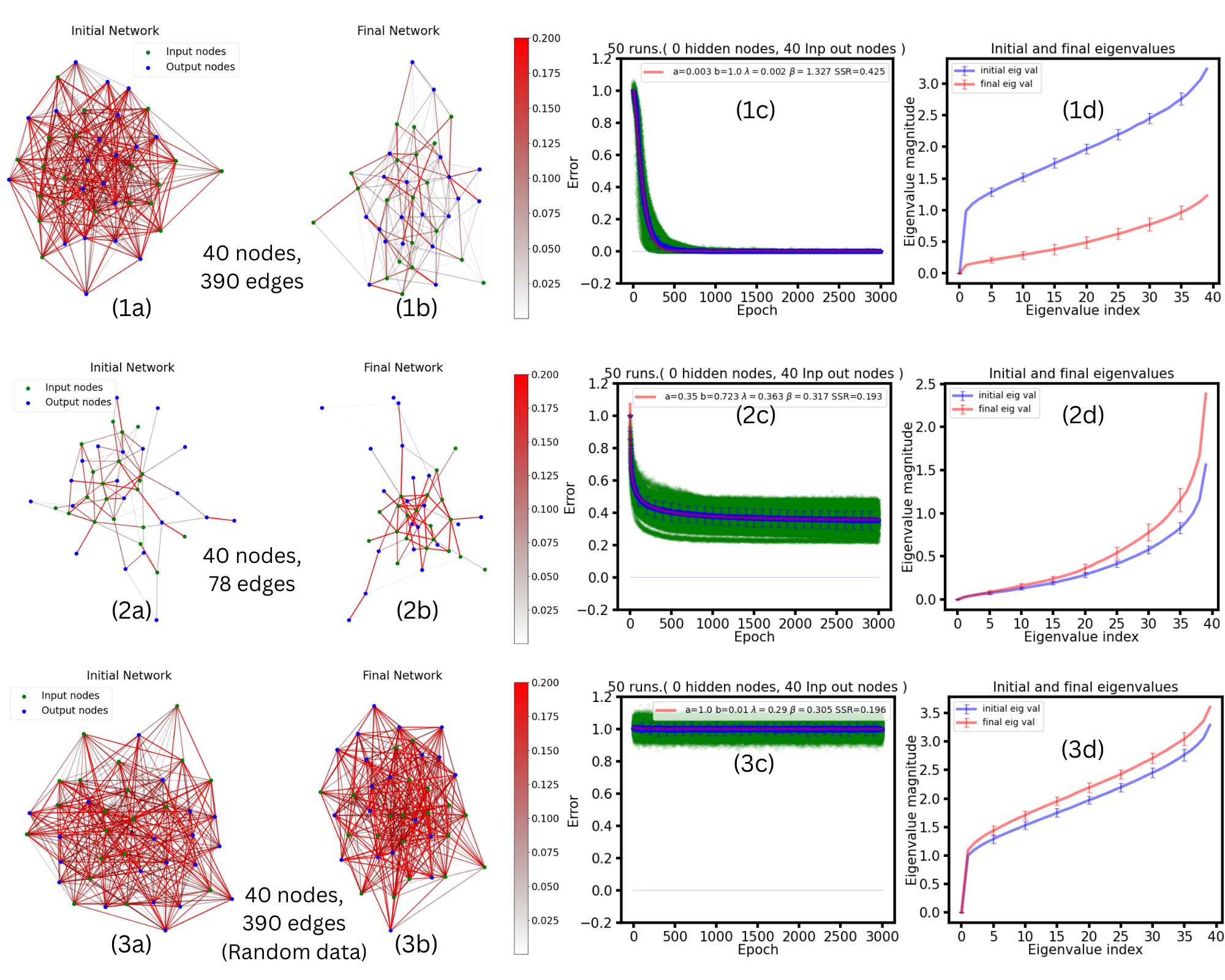}
    \caption{{\it Training linear networks to learn linear transformations.} [1a \& 1b] : {\it Network undergoes trimming}. A network with 40 nodes and 390 edges is trained to learn a linear transformation of size $10 \times 10 $. Weights of the network are uniformly sampled from $[10^{-5},0.2]$. Colorbar on right shows weight values of each edge. [1c]  {\it Non-exponential relaxation} : Training curve for the case shown in 1a and 1b but for 50 different initializations (shown in green). Y axis shows error defined as square root of mean square error, X axis shows epoch. In one epoch the network goes through 100 data points . All green curves are obtained after normalization with their respective inital errors. The blue curve shows the average over these 50 runs. The blue curve is fit to a non-exponential curve of the form $a+b e^{- \lambda \cdot t^{\beta}}$. Fit parameters are shown in the legend. $\beta>1$ shows the relaxation shows a compressed exponential behaviour. The sum of squared residuals (SSR) is used to assess the goodness of fit, it is defined as: $\text{SSR}=\sum_{i=1}^{n} (y^{fit}_i - y^{data}_i)^2 $ [1d] {\it Eigenvalues decrease while learning}: Eigenvalues of graph Laplacian before and after training for runs shown in 1c. These initial and final eigenvalues are averaged over those 50 runs. The eigenvalues are sorted in increasing order. The x-axis shows eigenvalue index. The network has 40 nodes so there are 40 eigenvalues. [2a, to d] These plots show the training performance for a network with less number of edges (78 edges), due to which it does not learn well. When compared with case 1, we see that trimming is less prominent and the eigenvalues do not decrease. The training curve shows a stretched exponential relaxation ($\beta<1$) and saturates well above zero error. [3a, to d] {\it Training on random data}: Networks initialized with same parameters as that of 1a are trained on randomly generated data. No trimming is observed, eigenvalues increase over training and the error curve does not decrease with the number of epochs. }  
    \label{fig:1}
  \end{figure*}

\section{The learning process}
\label{sec:the-learning-process}
We now demonstrate the process of physical learning within our system. Initially, we impose an input boundary condition, denoted by $I$. The system's response is then captured by the Laplace equation $Lv=I$, where $L$ is Laplacian, which depends on the learning degrees of freedom $w$,  and $v$ is the state of the system.
To attain its intended functionality, the system need to update $w$ to minimize the cost function $C(v(w))$. We encode the cost function as an interaction energy between the system and the environment. This energy causes a feedback boundary condition of the form $- \eta \dfrac{\partial C(v)}{\partial v}$ to act on the system, due to which the state of the system evolves along a direction that decreases $C(v)$: 
\begin{align}
        L(v+\delta v) = I - \eta \dfrac{\partial C(v)}{\partial v}.
        \label{state-shift}
\end{align} 
For a mechanical network, these input and feedback boundary conditions are applied as external stresses on the system. When the feedback stress is removed, the system tends to revert to its initial state $v$. However, with continuous exposure to feedback boundary forces, there's a lasting change in the system's learning degrees of freedom. This change is akin to a plastic deformation in materials where repeated stress leads to permanent alterations.

Note that unlike the input boundary condition, the feedback boundary condition is a function of the state of the system. As a result, there exists an optimal state where the system experiences minimal feedback stress. Our hypothesis is that, through repeated application of these feedback stresses, the system's learning parameters $w$ evolve such that this optimal state is reached. The objective of this evolution is to minimize the external stress $- \eta \dfrac{\partial C(v)}{\partial v}$, by changes in state of the system $v$, through changes in $w$. This adaptation is represented as:
\begin{equation}
        \Delta w_{ij} = -\alpha \eta \dfrac{\partial C(w)}{\partial w_{ij}}, 
        \label{eq:weight_update_glass}
    \end{equation}
where $C$ is a function of $w$ via $C(v(w))$. In our previous work~\cite{Anisetti2023}, we showed that the above weight update rule can be written purely in terms of local physical quantities\begin{equation}
    \Delta w_{ij}=-\alpha  v_{ij}  \delta v _{ij}. 
    \label{eq:weight_update_local}
  \end{equation} Where, $w_{ij}$ is the weight connecting nodes $(i,j)$, and $v _{ij}$ is the potential drop $v_{i}-v_{j}$, $\delta v _{ij}$ is the change in this potential drop due to feedback\footnote{
  We could also have defined $v _{ij}=v_{j}-v_{i}$, note that the learning rule is independent of this choice }. Intriguingly, this learning rule exhibits a Hebbian-like behavior.
  
The input and feedback boundary conditions encode a particular-type of information, and given that they are applied repeatedly, a parallel with memory formation in driven disordered systems seems plausible~\cite{keim2019memory}. For example, in granular systems, the particles rearrange in response to a particular sequence of driving amplitudes\cite{Corte2008,keim2011generic}. Additionally, if the network topology is fixed, then the learning degrees of freedom are updated much like in the context of directed aging~\cite{pashine2019,hexner2020effect}, keeping in mind that the update rule in our system depends on both feedforward signal ($v_{ij}$) and feedback signal ($\delta v _{ij}$) , rather than a reduction of spring constants over time based on the stress experienced by a particular spring. \\ \\
Due to the evolution of the learning degrees of freedom, once reaching steady state, the system's response is :\begin{equation}
        L'(v+\delta v) = I,
    \end{equation}
   where $L'$ is the updated Laplacian that encodes the memory of the feedback stress by adapting to it, i.e; $C(v+\delta v) < C(v) $.

 In summary, the learning process goes as follows. An input is introduced to the system as an external force. Subsequently, based on the system's reaction to this input, feedback forces are consistently applied. We postulate that such a process enables the system to adapt and become attuned to these feedback boundary forces. This continuous adaptation to feedback forces, in presence of the input, ingrains a memory of the input-output relationship within the system. This concept is elucidated further in the subsequent section.

\section{A brief review of Multi-mechanism Learning}
\label{sec:theory}
We study a network comprised of nodes and connected by weighted edges. Let us represent the weight of the edge between node \(x\) and node \(y\) as \(w_{xy}\), which could signify conductances in an electrical network, spring constants in a mechanical spring network, or pipe thickness in 
a flow network, etc.\\
\textbf{Input Nodes:} An ``input'' node pair is pair of nodes  $(b_j^+,b_j^-)$ such that an input current \(I_j\) enters the network via node \(b_j^+\) and exits through \(b_j^-\).(For mechanical networks input current can be thought of as external forces acting at input nodes). Let there be $q$ such input node pairs in the network, denoted by \(\{(b_1^+,b_1^-), (b_2^+,b_2^-), \ldots, (b_q^+,b_q^-)\}\).\\
\textbf{Output Nodes:} In response to the input currents, the system develops an electric potential at each node. The network's output is defined to be the set of potential differences across certain ``output'' node pairs, obtained as \(v(o_i^+,o_i^-) = v(o_i^+)-v(o_i^-)\) for each output node pair (\(o_i^+,o_i^-)\). Let there be p such output node pairs in the network, represented as \(\{(o_1^+,o_1^-), (o_2^+,o_2^-), \ldots, (o_p^+,o_p^-)\}\). \\
\textbf{Cost Function:} The goal of training is to adjust the weights \(\{w_{xy}\}\) so that for a given set of input currents, the desired potential drops \(\{v_d(o_i^+,o_i^-)\}\) are achieved across all the output nodes. We employ a Mean Squared Error (MSE) cost function:
\begin{equation}
    \label{eq:cost-function}
    C = \frac{1}{2} \sum_{i=1}^{p} (v(o_i^+,o_i^-)-v_d(o_i^+,o_i^-))^2.
\end{equation}\textbf{Feedback Mechanism:} To optimize this cost function, we introduce a feedback signal into the network at the output nodes. For each output node pair, the feed-back current is calculated as:
\begin{equation}
    \label{eq:chemical-current}
    \epsilon_i = -\eta (v(o_i^+,o_i^-)-v_d(o_i^+,o_i^-))
\end{equation}
This current enters the network through node \(o_i^+\) and exits via \(o_i^-\), with \(\eta\) being a positive ``nudging'' factor. The feedback currents change the potentials at each node and let the change in the potential at node $j$ be denoted by $u_j$. \\
\textbf{Weight Update Rule:} The weights are then updated as:
\begin{equation}
    \label{eq:learning-rule}
    \Delta w_{xy} = - \alpha u(x,y)v(x,y),
\end{equation}
where \(\alpha\) is the learning rate. This rule effectively performs gradient descent on the cost function:
\begin{equation}
    \label{eq:gradient-descent}
    \Delta w_{xy} = - \alpha \eta \dfrac{\partial C}{\partial w_{xy}}
\end{equation}
\textbf{Considerations:} The weight update is local, and its sign depends on the potential drops due to input and feedback. We assume the system's relaxation time is much shorter than the weight update time, ensuring a steady state during weight adjustments. The two quantities in the weight update must be independent. This can be ensured by encoding them into distinct physical quantities\cite{anisetti2022frequency}. (Further details on the learning procedure and its physical implementation are given in Ref.\cite{Anisetti2023}). For larger networks, a higher learning rate is necessary to maintain the magnitude of weight changes. To address this, we conduct a trial run for one epoch, adjusting the learning rate to ensure \(||\Delta w|| \approx 10^{-3}\). Additionally, we impose regularization by 
(1) Limiting each weight update: \( |\Delta w_{xy}| < \epsilon \) ,and (2)  Constraining weight values: \( w_{min} \leq w_{xy} \leq w_{max} \). This ensures a smooth training process and prevents weights from becoming too large or too small. In our simulations, we set \(w_{min}=0.00001\), \(w_{max}=0.2\), and \(\epsilon=0.01\).
\begin{figure*}
    \centering
    \includegraphics[width=16cm]{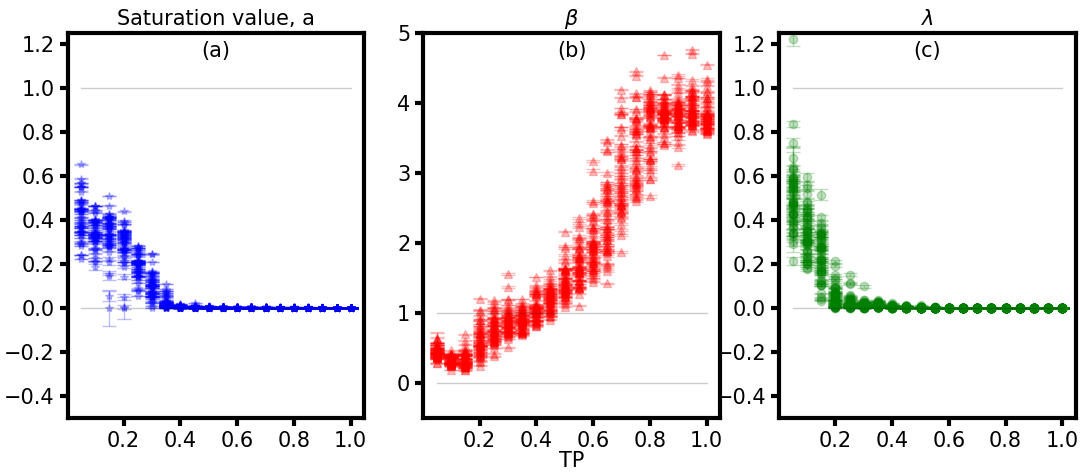}
    \caption{\textit{Learning performance with overparametrization}. Error curve is fit to $a+b e^{- \lambda\, t^{\beta}}$ for networks with varying edges and the fit parameters are plotted (Error bars shown are calculated using the diagonal terms of covariance matrix). The Tuning Parameter ($TP$) serves as a metric to quantify the degree of connectivity in a network. Specifically, it is calculated by taking the ratio of the number of edges $M$ present in the graph to the number of edges that would exist in a fully connected network with the same number of nodes.  (a) We observe that after adding a certain number of edges, the saturation value of the error curve begins to asymptote to zero. (b) We also observe that the exponent $\beta$ increases from less than one to greater than one, showing a shift from stretched exponential to compressed exponential relaxation. (c) $\lambda$ value also becomes very small after adding a certain number of edges. We have done a fit robustness analysis for these plots in Appendix A. (In Fig.\ref{fig:1}, 390 and 78 edge networks correspond to a $TP$ of 0.5 and 0.1, respectively.) }   
    \label{fig:2}
\end{figure*}

\section{Methodology}

{\bf Network Generation}: We aim to create networks consisting of $N$ nodes, with a varying number of edges $M$. For this, we first create a Barabási-Albert network with connection parameter 1. This graph generation algorithm connects a new node with 1 existing node in a manner that nodes with higher degree have a stronger likelihood for selection.This creates a network with $N$ nodes and $N-1$ edges. To create a network with $M$ edges, we add $M-(N+1)$ unique edges. This way, we can create networks with varying connectivity, ranging from being minimally connected to being maximally connected. Note that it is highly unlikely to create such minimally connected networks using the Erdős–Rényi model.

The generated networks are then trained on data generated using a linear transformation.  Note that in spite of using linear networks to learn linear transformations, the optimization needs to take place in a cost landscape which is non-convex, high- dimensional, and disordered.

{\bf Data Generation}: The input vector $\textbf{x}$ (eg; $(x_1,x_2,x_3)$) is encoded as external currents across input nodes
$\{(b_1^+,b_1^-), (b_2^+,b_2^-), (b_3^+,b_3^-)\}$ with currents $+x_q$ and $-x_q$ applied
across nodes $b_q^+$  and $b_q^-$ respectively. The output vector $\textbf{y}$ (eg; $(y_1,y_2,y_3)$ ) is the potential drop across nodes $\{(o_1^+,o_1^-), (o_2^+,o_2^-), (o_3^+,o_3^-)\}$. When the network is trained we want the network's output to closely approximate the  matrix $R$, that is we want $ \textbf{y} \approx R\textbf{x}$. To do so, we first generate training data of the form $\{(\textbf{x},R\textbf{x}) \}$ by randomly sampling $\textbf{x}$ from the surface of a unit sphere, and train the network using the procedure described in the previous section. To shorten the training time, we want the magnitude of output $\textbf{y}$ to be of the same order as that of the input, therefore we make sure that the maximum eigenvalue of $R$ is close to one. We do this by first generating an arbitrary matrix $R^{\prime}$ with random entries between -1 and 1, and then normalizing it by dividing it with maximum eigenvalue : $R=R^{\prime}/ max\{ eig(R^{\prime})\}$. Input and output data is generated using this matrix $R$. The network is trained using this ideal data, meaning each training step sees an entirely new data point. In the computer science community, this type of task is known as linear regression.

\begin{figure*}
    \centering
    \includegraphics[width=12cm]{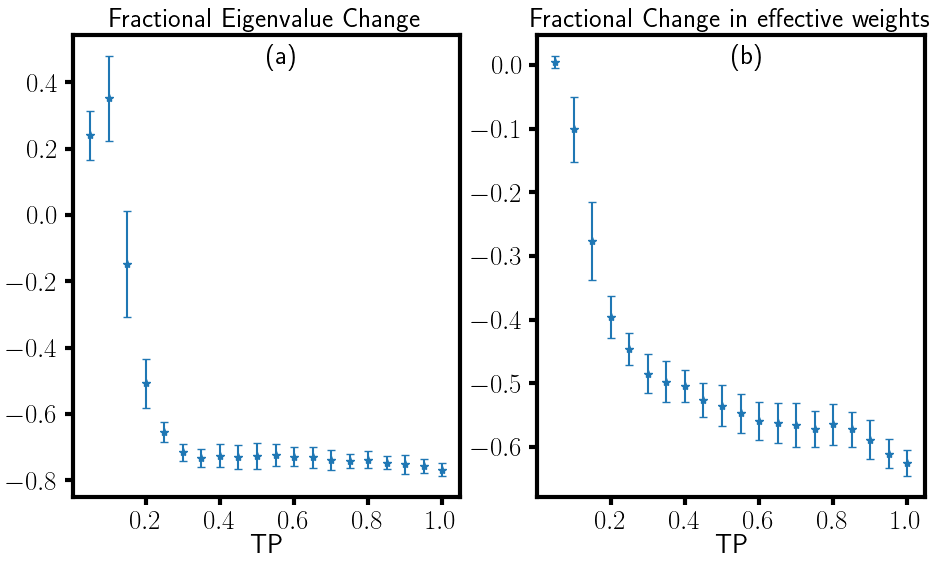}
    \caption{\textit{ Eigenvalue decrease and trimming with overparametrization }. (a) Shows fractional decrease in the sum of eigenvalues due to learning, averaged over 50 runs. (b) Shows fractional decrease in number of effective weights due to learning, averaged over 50 runs. Here, the term `effective weights' refers to those weights that fall within the top 99 percent of the permissible weight value range($[10^{-5},0.2]$). }   
    \label{fig:3}
\end{figure*}

\section{Results}
 
Figure~\ref{fig:1}.1a,1b shows the network before and after training for a network of $N=40$ and $M=390$. Since the intensity of the color indicates the magnitude of the weight, note that many of the weights of a trained network reach the minimum value. In other words, there is a trimming effect, where only the important edges remain. To ascertain whether or not the network has learned the linear transformation, we plot the square root of the mean-squared error in Fig. \ref{fig:1}.1c as a function of epoch. Given that the error nearly vanishes at longer epoch, this network has successfully learned the task. This shows the dynamics through which the system relaxes to the feedback boundary forces due to the evolution of learning degrees of freedom. Interestingly, we performed a phenomenological fit for this curve. The curve is well-approximated by a non-exponential relaxation of the form $\sqrt{MSE}=a+b\exp(-\lambda\,t^\beta)$, where $a$, $b$, $\lambda$, $\beta$ are the fit parameters and $t$ denotes the epoch number. Interestingly, these dynamics are quantitatively similar to what is observed in molecular glassy systems~\cite{Phillips1996}. This finding demonstrates the existence of a glassy landscape. Appendix A addresses the reasonableness of this non-exponential fit. 

We seek to quantify further the relaxation of the system as it learns. We, therefore, compute the eigenvalues of the Laplacian matrix. Figure\ref{fig:1}.1d shows how learning results in decreasing Laplacian eigenvalues. Note that these eigenvalues are the square root of normal mode frequencies. Decreasing eigenvalues is evidence that the network is getting ``softer" as the normal mode excitations become longer in wavelength. This observation demonstrates that the network moves from a state of stress to that of less stress due to repeated application  feedback boundary forces. The network is, thus, ``adapting" to these feedback forces indicating a transition towards a state that encodes a memory of the input-output relationship. Additionally, it draws parallels between this behavior and the self-organization observed in periodically sheared suspensions, where the system adapts to the periodic driving in a similar manner~\cite{Corte2008}. Moreover, when amorphous solids, modeled as purely repulsive particles in the jammed phase, are shear-stabilized by minimizing the energy with respect to the shear degrees of freedom, one finds longer wavelength excitations emerging~\cite{mizuno2017continuum}. Finally, recent work demonstrates that using a similar multiplicative learning rule as given in Eq. 7 to train physical networks to learn linear transformations also shows a decrease in the lowest eigenvalues of the Hessian~\cite{Stern2023physical}. Appendix \ref{Appendix:B} shows that the trends hold for larger system sizes. 

Figures 1.2(a-d) show the same quantities as Figure 1.1, however, for a network with $N=40$ and $M=78$. Given the smaller number of learning degrees of freedom, a network with this architecture does not successfully learn, as indicated by the square root of the mean-squared error not decreasing to zero as the number of epochs increase. Moreover, the eigenvalues of the Laplacian do not decrease and so the system does not relax, or soften. For comparative purposes, we also train the network to learn, if you will, random data.  Fig. \ref{fig:1}.(3a to d) shows the physical effects of learning random data. Here, the system, exposed to random input and feedback boundary conditions, does not relax, as indicated by the unchanged initial and final eigenvalues. With random input-output forces, the weight update signal in Eq. \ref{eq:weight_update_local} averages to zero due to the absence of correlation between $v_{ij}$ and $\delta v_{ij}$. This null result suggests that the system's relaxation is driven by correlations between input and feedback boundary conditions and for certain network architectures. 

Given the nontrivial dependence of learning on the network architecture, we further extend our analysis by incrementally increasing the network connectivity to examine the implications of overparametrization (see Fig. \ref{fig:2}). We denote the ratio of the number of edges $M$ to the number of edges in the fully connected equivalent network as $TP$ for tuning parameter. The results indicate that as more edges are introduced, the cost landscape becomes steeper due to a reduced number of flat directions~\cite{BaityJesi2019}, leading to accelerated relaxation and enhanced learning performance. Notably, a parallel can be drawn with glasses; in these systems, increased connectivity also speeds up relaxation dynamics~\cite{cui2016relation,Wu2018}. Both these studies, as well as ours, show a shift in relaxation dynamics from a stretched to a compressed exponential upon increasing connectivity. This further underscores the intrinsic link between learning processes and relaxation in disordered systems.

Given the changes in the weights as the networks learns, in Fig. \ref{fig:3}, we examine the relationship between trimming, eigenvalue reduction, and network connectivity. As network connectivity increases by increasing $TP$, the fractional eigenvalue decrease tends to plateau, reaching a saturation point around $TP$ $\approx 0.3$. A comparison of Fig. \ref{fig:3}(a) and Fig. \ref{fig:2}(a) reveals a notable correlation: the point of eigenvalue saturation aligns with the disappearance of saturation error. This suggests a fundamental link between the processes of learning and eigenvalue reduction. Furthermore, Fig. \ref{fig:3}(b) underscores the ubiquity of the trimming effect across networks of varying connectivity. Notably, the magnitude of the trimming effect intensifies as network size grows.

\begin{figure*}
    \centering
    \includegraphics[width=12cm]{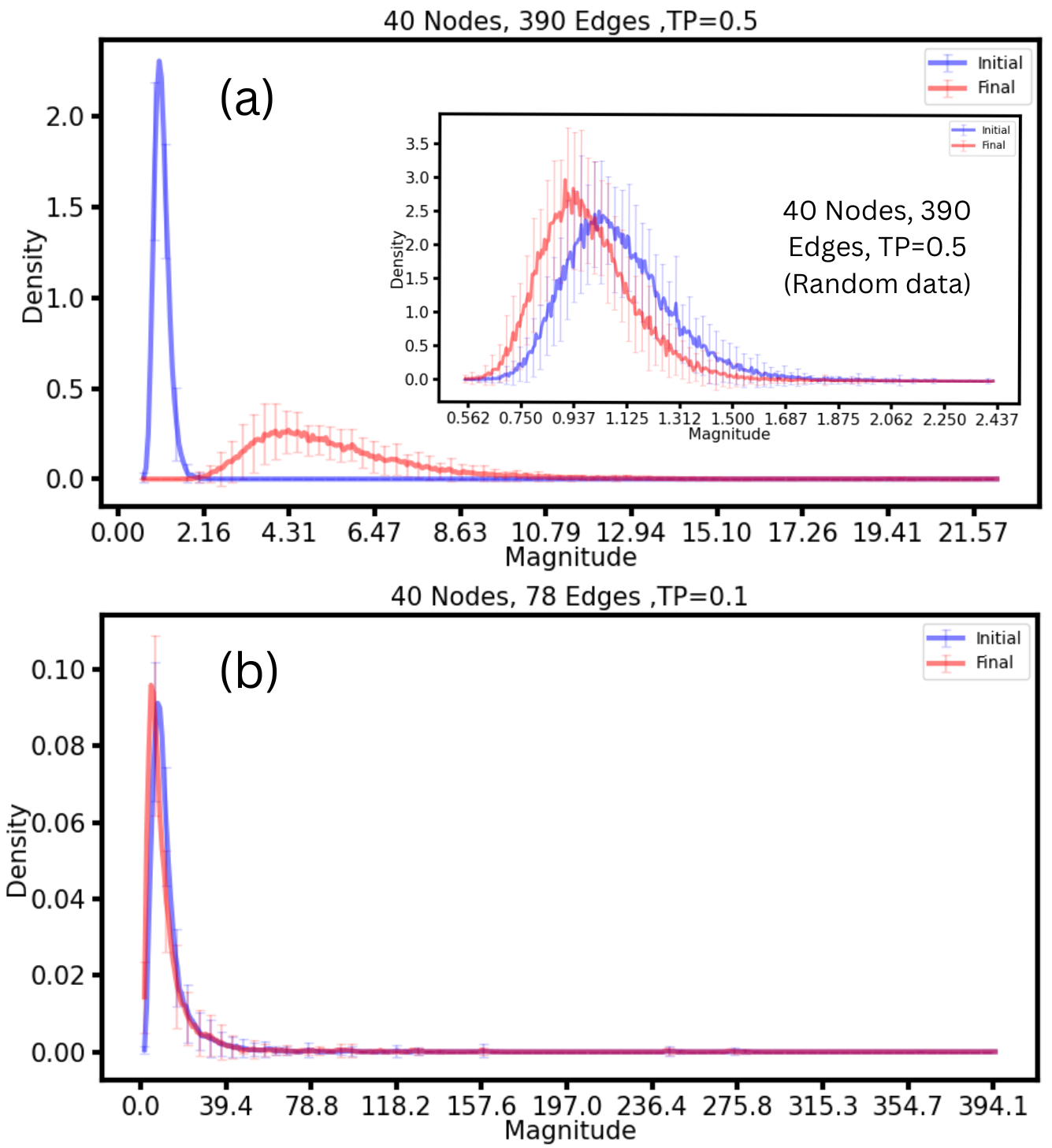}
    \caption{\textit{Resistance Distance Distribution and Learning}. (a) The figure showcases the average resistance distance distribution, $p(r)$, during learning, with the x-axis denoting resistance magnitude and the y-axis its normalized frequency. This is averaged over 50 network initializations. The inset illustrates the outcome when the network is trained on random data (note that the scale in the inset differs, making the initial distributions appear distinct, though they are identical). (b) Represents a network with suboptimal learning performance due to a limited number of edges.} 
    \label{fig:4}
\end{figure*}

Figure \ref{fig:4} illustrates the evolution of the resistance distance distribution during the learning process. In an electrical network, the effective resistance between two nodes can be interpreted as a measure of distance (more details in Appendix C). By calculating the average distribution of resistance distances over all possible pairs of nodes, a two-point correlation function $p(r)$ can be derived, which can be extended to spring and flow networks as well. As learning progresses, we observe a broadening of the two-point correlation function, indicating that the average conductance between two arbitrary nodes decreases. This phenomenon is analogous to a reduction in ``stiffness" in elastic networks, as the system becomes more soft during learning.

\section{Discussion}

In summary, in learning about the physical signatures of multi-mechanism learning we find that: 
\begin{enumerate}
    \item The error curve for networks with low connectivity resembles a stretched exponential. However, as network connectivity increases, the error curve transitions to a compressed exponential form (Fig. \ref{fig:2}).
    \item Eigenvalues of the graph Laplacian decrease with epoch and long wavelength modes are generated (Fig. \ref{fig:1}).
    \item The network undergoes trimming, i.e., lot of the weights go to zero (Fig. \ref{fig:1} \& \ref{fig:3}).
    \item The two point correlation function for the network broadens while learning (Fig. \ref{fig:4}).
\end{enumerate}
The non-exponential relaxation indicates the presence of a glassy learning landscape. In such a landscape, many local minima exist thereby allowing multiple memories to form. Interestingly, in prior work the optimization landscapes of Deep Neural Networks (DNNs) were compared with those of glasses~\cite{BaityJesi2019,Spigler2019}. A fundamental distinction was observed: while glasses exhibited slow relaxation over extended timescales, DNNs did not manifest such slow relaxation at long times. This discrepancy was hypothesized to arise from the overparametrization inherent to DNNs. Our findings, as illustrated in Fig.~\ref{fig:3}, corroborate this hypothesis. We demonstrate that even in physically disordered systems, increasing network connectivity can eliminate slow relaxation. This further suggests a potential SAT-UNSAT transition~\cite{monasson2007introduction} in these physical learning systems.

Experiments on directed aging~\cite{pashine2019} reveal that materials subjected to external stress undergo alterations in their physical properties and by meticulously controlling the application of this external stress, one can tailor a material to exhibit specific desired properties. A trivial example of this principle in action can be observed with a simple piece of paper. If one aims to create a material with a negative Poisson's ratio, the paper can be crumpled into a ball. When this crumpled paper ball is stretched horizontally, it also expands vertically for small strains, indicating a negative Poisson's ratio. To capture the essence of this behavior, previous studies have introduced a model where springs decrease their spring constants over time based on the local stress experienced by each spring~\cite{hexner2020effect,hexner2020periodic}. We posit that the model detailed in Section~\ref{sec:the-learning-process} offers a comprehensive explanation for this phenomenon. This is because directed aging, at its core, can be viewed as an adaptation to external stresses. Additionally, we believe that this approach can potentially explain adaptation to external driving that was observed in particulate systems, keeping in mind there are differences in memory formation between unjammed and jammed systems~\cite{Corte2008,keim2011generic,Keim2019,adhikari2018memory}.

Moreover, the softening of the system as it learns and the associated increase in of the correlation length suggest that the system is indeed relaxing into the imposed boundary conditions that encode the linear transformation. However, these boundary conditions contain more information than a simple scalar quantity such as a strain amplitude\cite{keim2019memory}. If the environment is simple enough, then the physical system can learn.  However, given too complex an environment, it may not be able to learn.  Of course, we have restricted ourselves to linear networks.  Nonlinear networks enhance the learning capability, as has been clearly shown in ANNs and even in mechanical networks~\cite{stern2020continual}.  
Neuromorphic researchers have been actively seeking physical counterparts to facilitate autonomous weight updates. This pursuit has led to the development of physical learning systems utilizing memristors~\cite{strukov2008missing}, nanoscale devices~\cite{kuzum2012nanoelectronic}, and transistors~\cite{jerry2017ferroelectric}. However, the intricate design requirements for each component presents challenges in terms of robustness and scalability. We propose that soft materials might offer a more streamlined solution. These materials inherently exhibit self-adjustment to external conditions, as evidenced by the self-organization of granular systems in response to external driving~\cite{Corte2008,keim2011generic,hexner2020periodic} the adaptability of other disordered systems to external strain~\cite{pashine2019,hexner2020effect}. Consequently, they emerge as promising candidates for crafting physical learning systems. Moreover, the model introduced in Section \ref{sec:theory} provides insights into a potential training methodology for soft materials, be it particulate-based, such as a granular learner, where the topology of the system can change, or spring-based, such a spring network learner, where the topology of the network is fixed. By iteratively applying input and feedback boundary forces, the learning parameters can autonomously adapt to these forces to optimize a cost function. This approach paves the way for the creation of innovative disordered materials with neural network-like learning potential. We aim to validate this concept in our forthcoming research.

Finally, by using multi-mechanism learning to train physical networks to learn linear transformations, we demonstrate a simple, brain-like task in a typically non-brain-like material.  As brains began to emerge several hundred million years ago in planarians~\cite{sarnat1985brain}, physical learning mechanisms are ripe candidates for life learning to survive in their environment before planarians. We, therefore, seek to validate such mechanisms in pre-planarian organisms.

The authors thank Benjamin Scellier, Arvind Murugan, Eli Hawkins, Shabeeb Ameen and Samuel Ropert for helpful discussion.  JMS acknowledges financial support from NSF-DMR-2204312.

\bibliography{learning} 
\bibliographystyle{ieeetr}

\setcounter{secnumdepth}{3}
\setcounter{equation}{0}
\setcounter{figure}{0}  
\setcounter{table}{0}

\makeatletter 
\renewcommand{\thefigure}{A\@arabic\c@figure}
\makeatother

\makeatletter 
\renewcommand{\thetable}{A\@Roman\c@table}
\makeatother

\makeatletter 
\renewcommand{\theequation}{A\@arabic\c@equation}
\makeatother
\appendix

\section{Analyzing the time dependence of the relaxation}
\label{Appendix:A}
\begin{figure*}
    \centering
    \includegraphics[width=15cm]{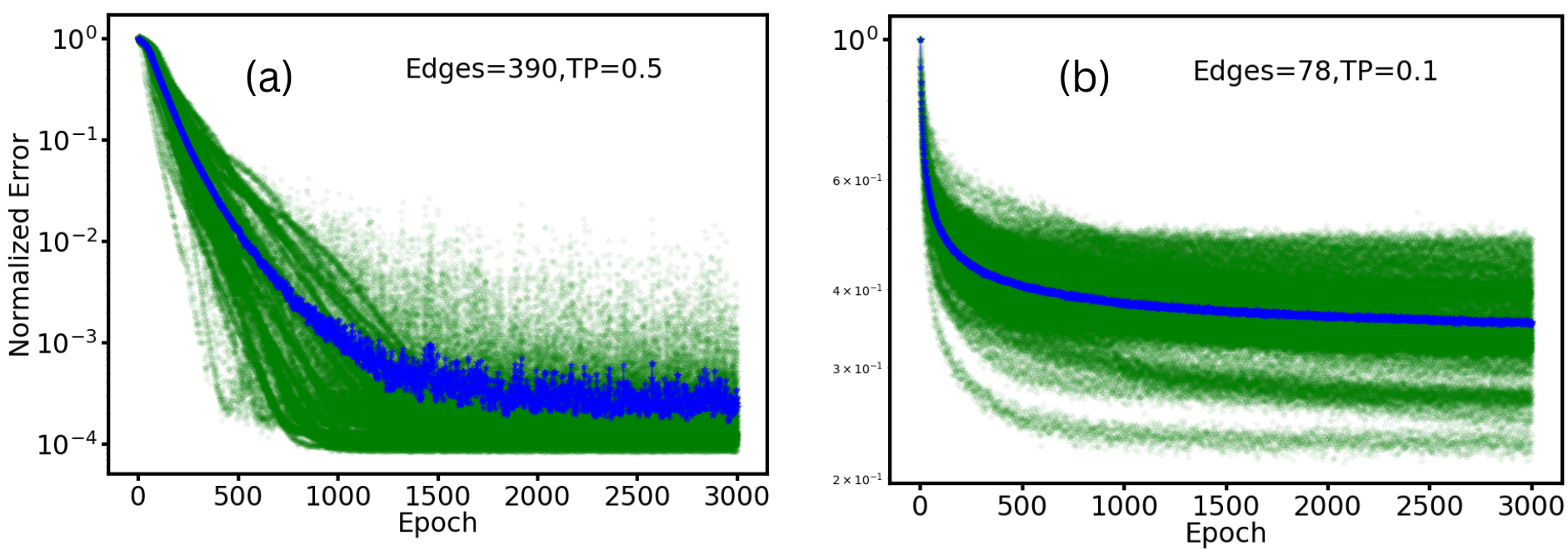}
    \caption{\textit{Log-Linear Analysis of Error Curves}: Panels (a) and (b) represent the log-linear plots corresponding to the error curves from Fig.~\ref{fig:1}:~2c and 1c, respectively. These analyses pertain to a 40-node network with 78 and 390 edges, respectively.}
\label{fig:7}
\end{figure*}

To more rigorously ascertain whether the error curves depicted in Fig.~\ref{fig:1} follow a non-exponential relaxation, we analyzed their log-linear plots. As evident in Fig.~\ref{fig:7}, these plots deviate significantly from a straight line, suggesting a departure from a simple exponential relationship.

\begin{figure*}
    \centering
    \includegraphics[width=15cm]{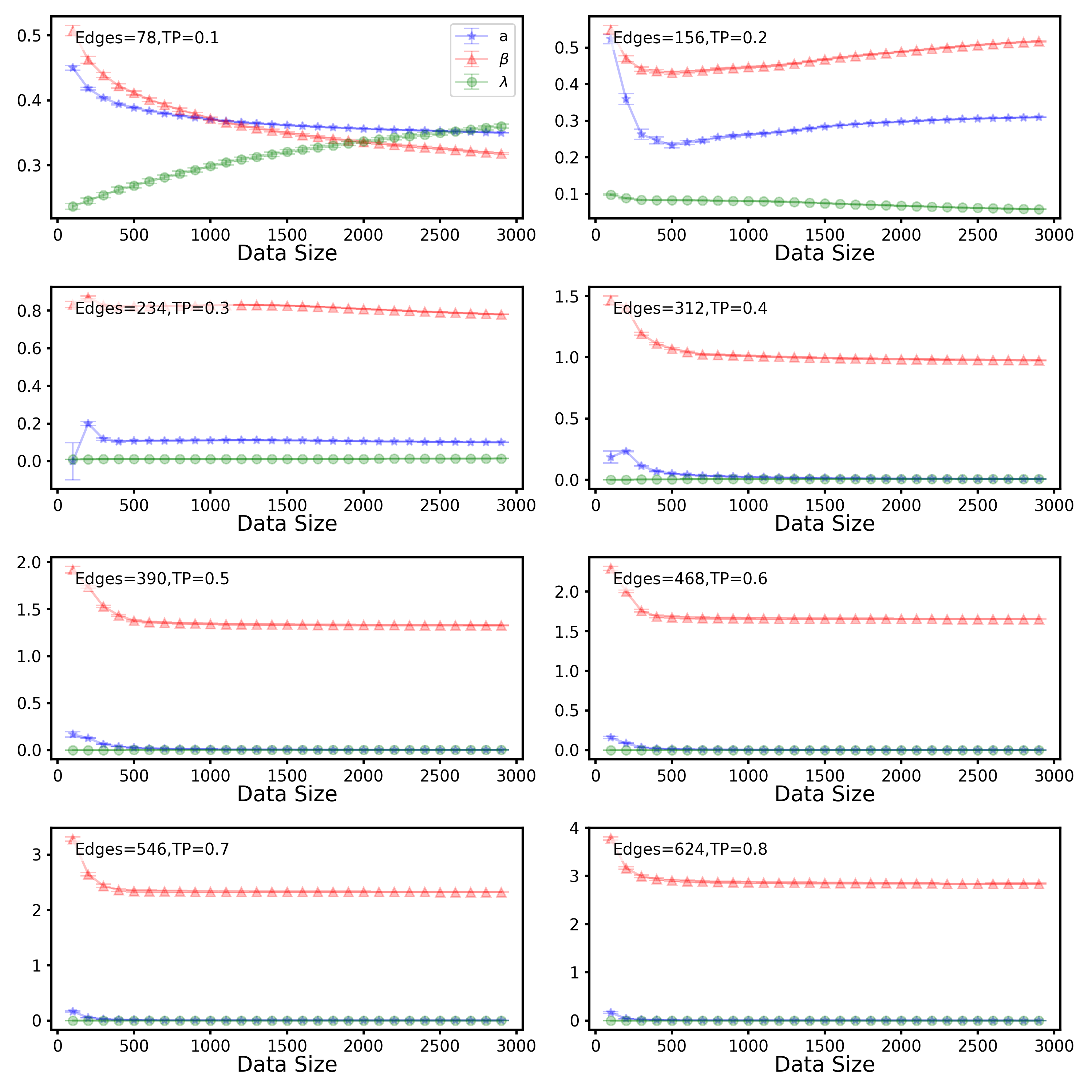}
    \caption{\textit{Fit Robustness Plots}: This figure illustrates the robustness of fit parameters depicted in Fig.~\ref{fig:2}. Each subplot represents the evolution of fit parameters with increasing data size. Notably, for $TP$ values of 0.1 and 0.2, the fit parameters exhibit less stability. In contrast, higher $TP$ values yield more consistent and stable parameters.}
\label{fig:5}
\end{figure*}

To demonstrate the significance of the fit parameters, we examined their behavior with progressive increments in the data size ( Fig.~\ref{fig:5} ), fitting them to the non-exponential function. We observed that increasing the data size leads the fit parameters to converge to a stable value, especially when the error curve reaches saturation. In some instances, the error curve takes an extended period to saturate, resulting in less stable fit parameters. This phenomenon is particularly noticeable for smaller $TP$ values, where the relaxation process is sluggish. Even with prolonged computations, the error curve does not achieve saturation under these conditions.
\section{Analysis for larger network sizes}
\label{Appendix:B}
\begin{figure*}
    \centering
    \includegraphics[width=15cm]{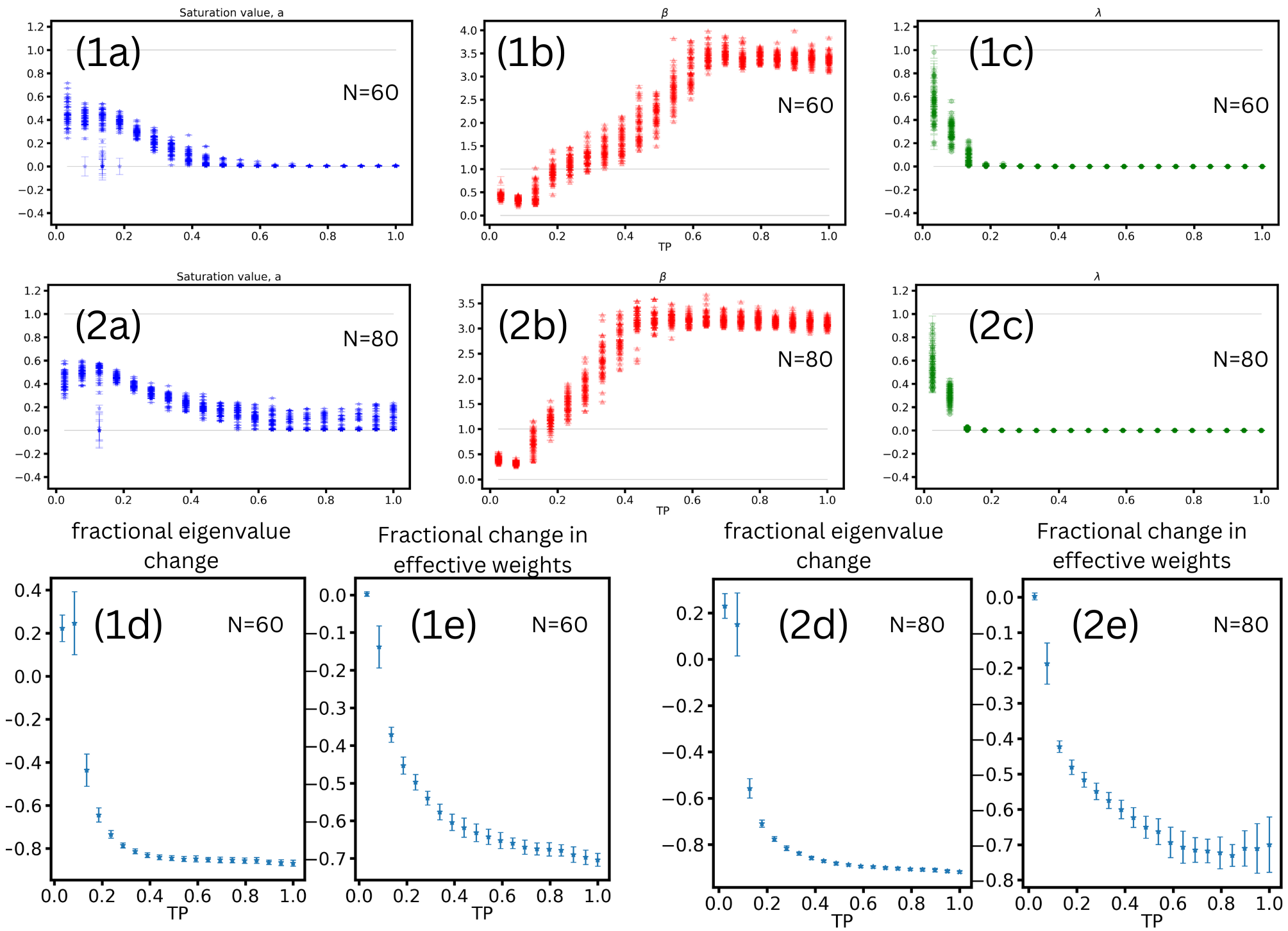}
    \caption{\textit{Analysis for larger networks}:[1a-c] shows fit parameters with increasing $TP$ and [1d-e] shows fractional eigenvalue change and fractional change in effective weights for 60 node networks [2a-e]. Shows similar analysis for 80 node networks.  }
\label{fig:6}
\end{figure*}

To further investigate the learning performance, we expanded our analysis from the previously studied 40-node network (as depicted in Fig.~\ref{fig:2} and Fig.~\ref{fig:3}) to networks consisting of 60 and 80 nodes, represented in Fig.~\ref{fig:6}. These networks were tasked with learning linear transformations of sizes \(15 \times 15\) and \(20 \times 20\), respectively, while varying the number of edges. We observe a sharper jump in $\beta$ value for larger network sizes, but the saturation value of the error curve starts saturating above zero.

\section{The two-point correlation function for resistance networks}
\label{Appendix:C}
The effective resistance between nodes $i$ and $j$ in a resistance network, denoted as $r_{ij}$, provides a measure of the `distance' between these nodes~\cite{spielman2023spectral_section12.8}. It quantifies the potential difference between nodes $i$ and $j$ when a unit current is injected at $i$ and extracted at $j$, normalized by the total current. The mathematical representation is:
\begin{equation}
r_{ij} = (\delta_i - \delta_j)^T L^+ (\delta_i - \delta_j) \end{equation} 
In this expression, $\delta_i$ and $\delta_j$ are the Kronecker delta vectors. For a network with $N$ nodes, $\delta_i$ is a vector of length $N$ that has a value of 1 at the $i$-th position and 0 everywhere else. Similarly, $\delta_j$ has a value of 1 at the $j$-th position and 0 elsewhere.
. The term $L^+$ represents the Moore-Penrose pseudoinverse of the Laplacian matrix $L$ of the network.

As we consider a network with a constant node count and progressively introduce more edges, the average `distance' between nodes diminishes. This is reflected in the reduction of the average path length, which represents the mean number of steps required to traverse from one node to another, averaged over all node pairs and paths. In the extreme case of a fully connected network, all nodes are adjacent. The effective resistance metric adeptly captures this behavior: as more edges are added, the effective resistance between nodes decreases.

Building on this understanding, we can conceptualize a two-point correlation function as: \begin{equation}
p(r) = \frac{1}{N} \sum_{i \neq j} \delta(r - r_{ij})
\end{equation}
Where $p(r)$ is normalized and N is chosen such that the area under p(r) is 1. In Fig.~\ref{fig:4} we plot this $<p(r)>$ averaged over 50 different network initializations.

\end{document}